\documentclass[useAMS,usenatbib]{mn2e}
\usepackage{graphicx}
\usepackage{amsmath}
\usepackage{amssymb}

\def\dark{1}
\def\leicester{2}
\def\nrao{3}

\title[Pre-Swift Radio Host Galaxies]{Late-Time VLA Reobservations Rule Out ULIRG-Like Host Galaxies For Most Pre-Swift Long-Duration Gamma-Ray Bursts}

\author[D. A. Perley et al.]
{Daniel A. Perley$^{\dark}$\thanks{e-mail: dperley@dark-cosmology.dk}, 
Jens Hjorth$^{\dark}$,
Nial R. Tanvir$^{\leicester}$,
Richard A. Perley$^{\nrao}$
\\
$^{\dark}${Dark Cosmology Centre, Niels Bohr Institute, University of Copenhagen, Juliane Maries Vej 30, 2100 K{\o}benhavn {\O}, Denmark} \\
$^{\leicester}${Department of Physics and Astronomy, University of Leicester, Leicester LE1 7RH, UK} \\
$^{\nrao}${National Radio Astronomy Observatory, P.O. Box O, Socorro, NM, 87801, USA} \\
}

\begin{document}

\date{}

\pagerange{\pageref{firstpage}--\pageref{lastpage}} \pubyear{2016}

\def\LaTeX{L\kern-.36em\raise.3ex\hbox{a}\kern-.15em
    T\kern-.1667em\lower.7ex\hbox{E}\kern-.125emX}

\newcommand{\citepeg}[1]{\citep[{e.g.,}][]{#1}}
\newcommand{\citepcf}[1]{\citep[{see}\phantom{}][]{#1}}
\newcommand{\rha}[0]{\rightarrow}
\def\etal{{\sl et al.}}
\def\lsim{\hbox{ \rlap{\raise 0.425ex\hbox{$<$}}\lower 0.65ex\hbox{$\sim$}}}
\def\gsim{\hbox{ \rlap{\raise 0.425ex\hbox{$>$}}\lower 0.65ex\hbox{$\sim$}}}
\def\arcmin{\hbox{$^\prime$}}
\def\arcsec{\hbox{$^{\prime\prime}$}}
\def\arcdeg{\mbox{$^\circ$}}
\def\fd{\hbox{$~\!\!^{\rm d}$}}
\def\fh{\hbox{$~\!\!^{\rm h}$}}
\def\fm{\hbox{$~\!\!^{\rm m}$}}
\def\fs{\hbox{$~\!\!^{\rm s}$}}
\def\ale{\mathrel{\hbox{\rlap{\hbox{\lower4pt\hbox{$\sim$}}}\hbox{$<$}}}}
\def\age{\mathrel{\hbox{\rlap{\hbox{\lower4pt\hbox{$\sim$}}}\hbox{$>$}}}}
\def\msyr{\hbox{M$_\odot$ yr$^{-1}$}}
\def\Swift{{\textit{Swift}}\,}
\def\Fermi{{\textit{Fermi}}\,}
\def\apjl{{ApJL}\,}
\def\apj{{ApJ}\,}
\def\nat{{Nature}\,}
\def\aap{{A\&A}\,}
\def\pasp{{PASP}\,}
\def\aj{{AJ}\,}
\def\procspie{{Proc.~SPIE}\,}
\def\mnras{{MNRAS}\,}
\def\apjs{{ApJS}\,}
\def\aaps{{A\&AS}\,}
\def\araa{{ARAA}\,}
\def\apss{{Ap\&SS}\,}
\def\physrep{{PhysRep}\,}
\def\href{{}}

\label{firstpage}
\label{lastpage}

\maketitle

\begin{abstract}
We present new Jansky Very Large Array observations of five pre-Swift gamma-ray bursts for which an ultraluminous (SFR $>100$ $M_\odot$ yr$^{-1}$) dusty host galaxy had previously been inferred from radio or submillimetre observations taken within a few years after the burst.  In four of the five cases we no longer detect any source at the host location to limits much fainter than the original observations, ruling out the existence of an ultraluminous galaxy hosting any of these GRBs.  We continue to detect a source at the position of GRB\,980703, but it is much fainter than it was a decade ago and the inferred radio star-formation rate ($\sim80 M_\odot$) is relatively modest.  The radio flattening at 200--1000 days observed in the light curve of this GRB may have been caused by a decelerating counterjet oriented 180 degrees away from the viewer, although an unjetted wind model can also explain the data.   Our results eliminate all well-established pre-Swift ULIRG hosts, and all cases for which an unobscured GRB was found in a galaxy dominated by heavily-obscured star-formation.  When GRBs do occur in ULIRGs the afterglow is almost always observed to be heavily obscured, consistent with the large dust opacities and high dust covering fractions characteristic of these systems.
\end{abstract}


\section{Introduction}
\label{sec:intro}

Long gamma-ray bursts are produced by the explosion of massive, short-lived stars at cosmological distances (\citealt{HjorthBloom2012}).  Their host-galaxy population should therefore reflect and reveal the diversity of star-forming galaxies responsible for the Universe's star-formation across cosmic history.  One type of galaxy we may expect to frequently observe GRBs originating from is the broad class of luminous, dusty star-forming galaxies (DSGs).  These include submillimetre galaxies (SMGs; galaxies at cosmological redshift detected at 850$\mu$m with single-dish telescopes), ultra-luminous infrared galaxies (ULIRGs; galaxies with infrared luminosity exceeding $>10^{12} L_\odot$), and similar systems containing extensive dust-obscured star-formation.  They are nearly absent in the low-redshift universe but are relatively common at $z>1$, where they play an important role in galaxy evolution and cosmic star formation (see \citealt{Casey+2014} for a review).  Large columns of interstellar dust obscure nearly all of the optical and UV light from young stars in galaxies of this type, making them difficult both to find and to study.  Observations at long wavelengths (mid-IR, submillimetre, and radio) where the dusty ISM becomes transparent are critical.

\begin{table*}
\begin{minipage}{130mm}
\caption{Previously Claimed Detections of ULIRGs Hosting Pre-Swift GRBs$^{a}$}
\begin{tabular}{lll|lll|lll|l}
\hline
{} &
{} &
{} &
\multicolumn{3}{|c|}{\underline{Radio}} &
\multicolumn{3}{|c|}{\underline{Submillimetre}} &
{} 
\\
{GRB} & {$z$} &  {OA?$^{b}$} & {Freq.} & {$F_\nu$ $^{c}$} & {Ref.$^{d}$} & {Freq.} & {$F_\nu$ $^{c}$} & {Ref.$^{d}$} &
{SFR$^{e}$} \\
{} & {} & {} & {(GHz)} & {($\mu$Jy)} & {} & {(GHz)} & {($\mu$Jy)} & {} & {($M_\odot$ yr$^{-1}$)} \\
\hline
980703   &  0.967  &yes (red) &   1.43 &      68 $\pm$ 7  & B01 & 350 &{\it $<$2280 }    & T04 & 180, 212 \\
         &         &          &   4.86 &      42 $\pm$ 9  & B01 &     &                  &     &     \\
         &         &          &   8.46 &      39 $\pm$ 5  & B01 &     &                  &     &     \\
\hline
000210   &  0.8452 &none (dark)&  8.46 &{\it  18 $\pm$ 9 }& B03 & 350 &    3050$\pm$760  & T04 & 560, 179  \\
\hline
000418   &  1.1185 &yes (red) &   1.43 &      59 $\pm$ 15 & B03 & 350 &    3150$\pm$900  & B03 & 690, 330, 288 \\
         &         &          &   4.86 &      46 $\pm$ 13 & B03 & 670 &    4199$\pm$1900 & B03 &  \\
         &         &          &   8.46 &      51 $\pm$ 12 & B03 &     &                  &     &  \\        
\hline
000911   &  1.0585 & yes      &   8.46 &{\it $<$40       }& B03 & 350 &{\it2310$\pm$910} & B03 & 495   \\
\hline
010222   &  1.478  & yes      &   4.86 &{\it  23 $\pm$ 8 }& B03 & 250 &    1050$\pm$220  & F02 & 610, 300, 278  \\
         &         &          &   8.46 &{\it  17 $\pm$ 6 }& B03 & 350 &    3740$\pm$530  & F02 &                \\
\hline
021211   &  1.006  & yes      &   1.4  &     330 $\pm$ 31 & M12 &     &                  &     & 825            \\
         &         &          &   2.1  &{\it $<$34       }& H12 &     &                  &     &                \\
\hline
\end{tabular}
$^{a}${\   We exclude GRBs 980329, 000301C, and 000926, which are listed as possible low-significance radio host detections by B03 but acknowledged to contain significant afterglow contribution.  We include 000911, which is not explicitly claimed as a detection by B03 but for which a submillimetre detection at $>2.5\sigma$ is presented in their plots and tables.} \\
$^{b}${\  Whether or not an optical afterglow was detected for this GRB.   Only GRB\,000210 was ``dark'', indicating that the GRB occurred in an optically-thick region.  GRBs\,980703 and 000418 show evidence for moderate ($A_V \sim 1-2$ mag rest-frame) extinction \citep{Klose+2000,Kann+2006}.  The remaining GRBs show no evidence for extinction within their host galaxies.} \\ 
$^{c}${\  Italicized for events for which the reported detection is less than 3$\sigma$ and for nondetections.} \\ 
$^{d}${\  References for reported flux.  B01 = \cite{Berger+2001}; F02 = \cite{Frail+2002}; B03 = \cite{Berger+2003};  T04 = \cite{Tanvir+2004}; M12 = \cite{Michalowski+2012b}; H12 = \cite{Hatsukade+2012}.} \\ 
$^{e}${\  Inferred submillimetre or radio star-formation rates from the referenced works and/or from \cite{Michalowski+2008}}
\end{minipage}
\label{tab:prevfluxes}
\end{table*}

The first luminous DSG candidates hosting GRBs were found incidentally: late-time flattenings of the light curves of GRB\,970803 and GRB\,010222 at radio and/or submillimetre wavelengths were interpreted as being due to host-galaxy emission (\citealt{Berger+2001,Frail+2002}).   A large amount of dedicated effort was also invested during the pre-Swift era in conducting late-time, long-wavelength observations specifically with the intent of looking for late-time host emission.   Some of these efforts \citep{Barnard+2003,Tanvir+2004} produced only upper limits.   However, the comprehensive survey of \cite{Berger+2003} produced radio detections of at least three (and possibly as many as seven, if marginal detections are considered) out of 17 GRB host galaxies observed with the Very Large Array (VLA) and Australian Telescope Compact Array (ATCA) in observations reaching flux limits of typically 30\,$\mu$Jy at 1.4--8 GHz (3$\sigma$), corresponding to star-formation rates (SFR) of few hundred $M_\odot$/yr at $z\sim1.5$.   \cite{Berger+2003} report a similar detection fraction at 850 $\mu$m (to limits of 3 mJy, or $\sim$500 $M_\odot$ at $z\sim1.5$).  

These observations were taken to support a simple picture, as follows.  First, in agreement with the consensus view, a significant minority of high-redshift star-formation occurred in very luminous DSGs.  Second, GRBs trace the global star-formation rate with reasonable fidelity (the fraction of stars that explode as GRBs is similar in DSGs and in other, more ordinary galaxies).

However, the reported properties of the DSGs hosting pre-Swift GRBs differ markedly from the properties of DSGs found by other means.   Classically-selected DSGs usually show some evidence of very active star-formation and dust extinction in the form of red optical/IR colors or exceptionally strong emission lines, and they usually have high stellar masses \citep{Michalowski+2012a}, often exceeding $>$10$^{11} M_\odot$.  Yet, despite truly tremendous submm/radio-inferred star-formation rates ($>$300--500 $M_\odot$\,yr$^{-1}$), many of the claimed pre-Swift submillimetre/radio hosts show blue colors, low apparent optical extinction, and low masses uncharacteristic of the SMGs found in submillimetre/radio surveys \citep{Michalowski+2008}.  Also, several were observed at 24 $\mu$m \citep{LeFloch+2006} and none of these were detected, even though 24$\mu$m observations are also thought to probe dust-obscured star-formation.

It is possible that the classical submillimetre field surveys were simply ``missing'' a large population of young submillimetre galaxies with blue colors, high temperatures, and strong silicate absorption at 24 $\mu$m \citep{Michalowski+2008}.  This would be an important result, since it would imply that a significant fraction of the Universe's stars formed in a class of galaxies that eludes classical surveys.

\begin{table*}
\begin{minipage}{150mm}
\caption{VLA Observations}
\label{tab:observations}
\begin{tabular}{llllllllll}
\hline
{GRB} &
{RA$^{a}$} & 
{Dec$^{a}$} & 
{$z$$^{b}$} & 
{Band} &
{Config.$^{c}$} &
{Observation date} &
{$t_{\rm int}$ $^{d}$} &
{Beam size$^{e}$} &
{RMS noise$^{f}$}
 \\
{} &
{} &
{} &
{} &
{} &
{} &
{(UT)} &
{(hr)} &
{($\arcsec$)} &
{($\mu$Jy/beam)}
\\
\hline
  980703  & 23:59:06.67  & +08:35:07.09 & 0.967  &C& C & 2014-10-18 & 1.02 & 4.6$\times$3.6\arcsec   & 2.9 \\
          &              &              &        &C& A & 2015-07-06 & 1.02 & 0.40$\times$0.33\arcsec & 3.1 \\
          &              &              &        &L& B & 2012-06-24 & 5.31 & 5.5$\times$5.5\arcsec   & 7.9 \\
  000418  & 12:25:19.3   & +20:06:11.6  & 1.1185 &C& C & 2014-10-17 & 1.04 & 3.8$\times$3.5\arcsec & 3.3 \\
  000911  & 02:18:34.36  & +07:44:27.7  & 1.0585 &C& C & 2014-11-20 & 1.26 & 4.3$\times$3.6\arcsec & 2.9 \\
  010222  & 14:52:12.55  & +43:01:06.2  & 1.478  &C& A & 2015-08-28 & 1.64 & 0.39$\times$0.34\arcsec & 2.6 \\
  021211  & 08:08:59.883 & +06:43:37.88 & 1.006  &C& C & 2014-10-22 & 0.79 & 4.3$\times$3.4\arcsec & 3.5 \\
\hline
\end{tabular}
$^{a}${\ Observation pointing centre (J2000).}
$^{b}${\ Redshift of host or afterglow.}
$^{c}${\ VLA array configuration.}
$^{d}${\ Total time on-source in hours, excluding overheads.}
$^{e}${\ Major and minor axis FWHM of the synthesized beam.}
$^{f}${\ Noise (1$\sigma$) estimated from the standard deviation of 1000 randomly chosen points in the final map.} 
\end{minipage}
\end{table*}

Curiously, however, few of the GRBs actually found within these DSGs were optically-obscured themselves (Table \ref{tab:prevfluxes}): the afterglows showed only modest or even no evidence for extinction, corresponding to $A_V \lesssim 1$ mag along the line of sight to the GRB in all but one case \citep{Kann+2006}.   It is hard to explain why, in a galaxy population purportedly dominated by optically-thick star-formation, the GRBs would occur in the bolometrically insignificant optically-thin regions.  While GRBs can destroy dust in their close vicinity ($\sim$10 pc; \citealt{Waxman+2000}, see \citealt{Morgan+2014} for a possible observed example), it is not likely that this is possible out to the more extended spatial scales relevant to DSGs.

Even more problematically, attempts to replicate the pre-Swift studies on the much larger Swift sample have not led to comparable success.  Large, deep radio and Herschel surveys have produced a few secure examples of DSGs with star-formation rates $>$100--300\,$M_\odot$ hosting GRBs \citep{Perley+2013b,Perley+2015,Hunt+2014,Schady+2014}. But none of these would have been detected to the shallower limits of pre-Swift observations---and the bursts hosting them were heavily obscured in almost all cases, even though both obscured and unobscured GRBs were searched.

It seems worth considering, therefore, that the pre-Swift long-wavelength late-time detections may not have been as robust as claimed a decade ago, or that they originated from some other process unrelated to star-formation in the host galaxy---in particular, afterglow emission.

In this paper we investigate this topic directly by testing whether the purported long-wavelength host galaxy emission reported in previous studies is still present a decade after the initial detections.  In \S \ref{sec:observations} we present new ultra-late-time ($>10$ years post-GRB) VLA observations of five proposed ULIRG-like submillimetre/radio-detected pre-Swift GRB host galaxies.  We detect none of the hosts at their previously-measured level.  Having ruled out a host-galaxy origin, in \S \ref{sec:results} we attempt to explain the previous data, and suggest that while some of these host ``detections'' were simply due to source confusion or statistical fluctuations, at least one provides evidence for interesting physical behavior of the afterglow on timescales of 1--5 years post-GRB: in particular, the possible emergence of the counterjet.  We conclude in \S \ref{sec:conclusions}.

\section{Observations}
\label{sec:observations}

The VLA underwent a significant upgrade in the late 2000s, improving the continuum sensitivity of the array by approximately an order of magnitude at most frequencies (the upgraded facility is referred to as the Jansky Very Large Array; \citealt{PerleyR+2011}).  Even short integrations with the upgraded array can provide images with an RMS sensitivity exceeding the deepest pre-Swift host observations in the literature by a significant margin.  Accordingly, we proposed for and obtained observations of all\footnote{Three DSG host candidates (GRBs 980329, 000926, 000301C) from \cite{Berger+2003} were not observed by our program, but in all of these cases the radio observations used to infer the presence of a host were at relatively early epochs when the afterglow contribution was known to be significant, and the reported detection of any host excess was less than $<3\sigma$.  Additionally, the host of GRB\,020819B has recently been shown to be at much higher redshift than previously assumed \citep{Perley+2016e}, indicating that this galaxy may be a DSG/ULIRG also.  However, it is not clear whether the putative radio afterglow (and therefore the host) is actually associated with the GRB.} well-established pre-Swift DSG-like host galaxies accessible to the VLA (Table \ref{tab:observations}).  Only one pre-Swift DSG host galaxy (GRB\,000210) was at a declination too far south for the VLA to observe.  It was recently observed (albeit to relatively shallow limits) with the ATCA by \cite{Greiner+2016}; we adopt the limiting flux from their paper in our analysis.  

All of our observations were conducted in C-band.  To maximize sensitivity and frequency coverage we employed the 3-bit samplers to cover nearly the entire receiver band, extending from 4--8 GHz with a central frequency at 6.0 GHz.  Most of our observations were conducted in the C-configuration (3.4 km maximum baseline), but GRB\,010222 was observed in A-configuration (36 km maximum baseline) instead, and GRB\,980703 was observed in both A-configuration and C-configuration.  Observations were typically 1.0--1.5 hour per field (switched several times with a nearby phase calibrator, and beginning or concluding with the observation of a standard flux calibrator).  The 1$\sigma$ RMS noise of the combined maps were typically 3 $\mu$Jy.

To supplement our own observations, we searched the VLA archive and found an additional unpublished observation of the position of GRB\,980703 in L-band (1.4 GHz), taken on 2012 June 24.  We downloaded these data and reduced the observation using similar techniques.

Data reduction was carried out using the Astronomical Image Processing System (AIPS).    Radio frequency interference was minor in all observations, and generally removed by clipping outlier visibilities above a minimum flux density threshold of five times the RMS noise.

The resolution of the C-configuration observations is sufficiently coarse ($3-4 \arcsec$ FWHM) that all of the host galaxies are expected to be much smaller than the synthesized beam and can be treated as point sources.  To measure the flux density of the target in these images, we simply take the measured flux of the VLA map at the location of the host.  In the case of the A-configuration-only observation of GRB\,010222 the beam size is small (0.37$\arcsec$ FWHM), and if the host were extended on a scale similar to or larger than this, our observations might resolve out extended structure and underestimate the total flux or flux limit.  However, Hubble Space Telescope imaging of this host shows it to be dominated by a compact core ($\sim$0.15\arcsec\ FWHM; \citealt{GCN1087}) that contributes most of the optical flux and is also where the GRB occurred---suggesting that the procedure adopted for the other images is valid for this observation as well.   We note, however, that our measurement applies only to the central starburst and a measurement/limit on the entire galaxy would be (slightly) higher.

\section{Results}
\label{sec:results}

\begin{figure*}
\centerline{
\includegraphics[width=6.6in,angle=0]{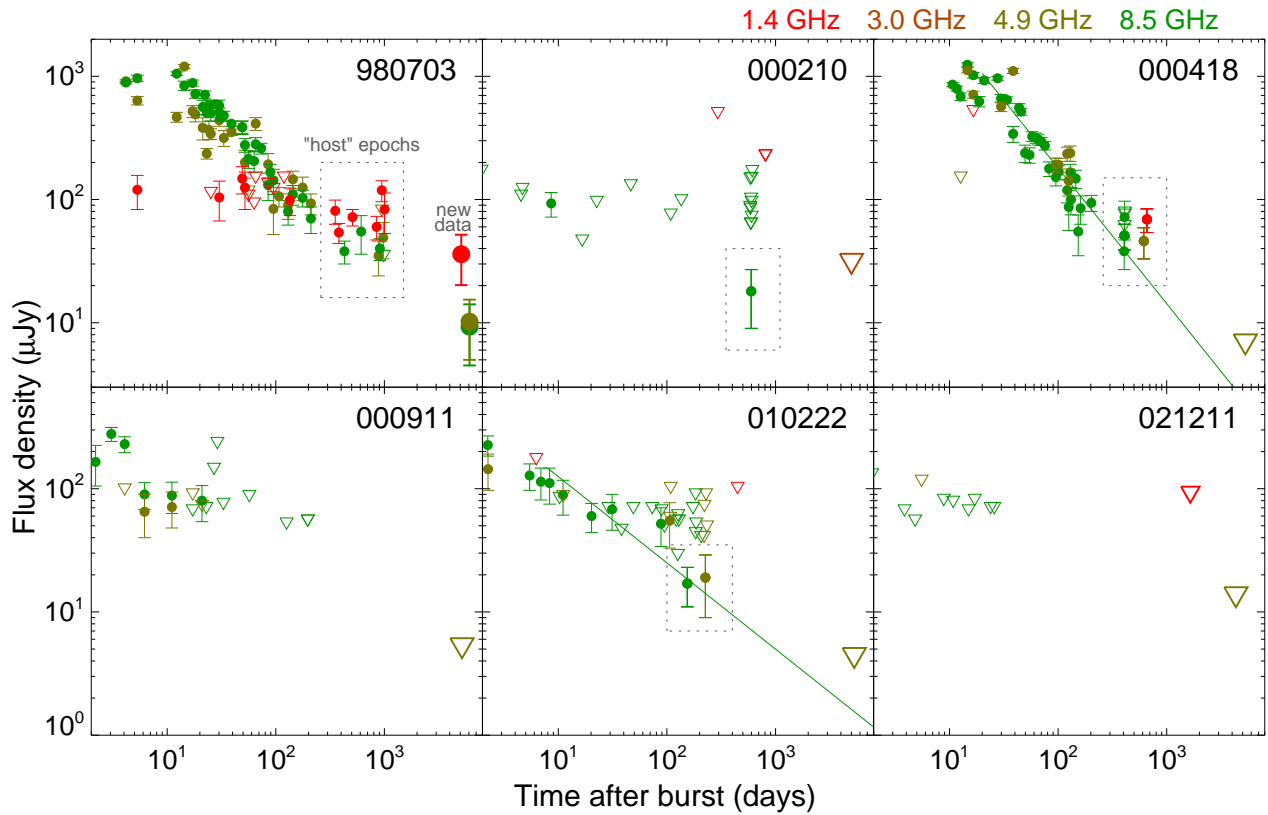}} 
\caption{Radio light curves of the six pre-Swift GRBs presented in Table 1.   Data are from Chandra \& Frail (2012) and from Berger et al.\ (2003); boxes are drawn around the putative host detections reported by Berger et al.   Our new ultra-late-time observations are also included, as well as the upper limit on GRB\,000210 from Greiner et al. 2016 (larger symbols).  Points are color-coded by (approximate) central frequency according to the legend at top right.  All sources previously suggested to represent the host galaxy have faded.  Most have disappeared below the detection threshold, with GRB 980703 representing the only exception.  In several cases the putative host detection was likely the result of late-time afterglow contributions: simple power-law extrapolation of the earlier observations (shown only for the 8 GHz data for clarity) are consistent with the late-time detections within 2$\sigma$.
}
\label{fig:radioag}
\end{figure*}

Only one of our targets is clearly detected in the new VLA imaging.  Directly at the afterglow (and host) position of GRB\,980703 we detect a source with a flux density of $10 \pm 2$ $\mu$Jy in the combined C-band data set.  A consistent flux density is measured from taking the C-configuration and A-configuration observations separately, suggesting that the source is compact.  Splitting the observations in frequency instead (but combining configurations/epochs), we measure flux densities of 10.2$\pm$2.6 $\mu$Jy (5 GHz), and 9.3$\pm$2.4 $\mu$Jy (7 GHz).  It is also detected in L-band; we measure 36$\pm$8 $\mu$Jy (1.4 GHz).

We marginally detect (2.0$\sigma$ significance) a weak source at the location of GRB\,021211, though its position is not exactly centred at the host location and it appears structured.  Likely it is a noise fluctuation.  None of the other sources show any significant flux excess at or near the position of the GRB.  Flux density measurements for all targets are summarized in Table \ref{tab:fluxes}.

\begin{table}
\begin{minipage}{80mm}
\caption{Host-Galaxy Flux Densities and Star-Formation Rates}
\label{tab:fluxes}
\begin{tabular}{lllll}
\hline
{GRB} & {$z$} & {Frequency} & {Flux density} & {SFR}  \\ 
{} & {} & {(GHz)} & {($\mu$Jy)} & {($M_\odot$ yr$^{-1}$)} \\ 
\hline
980703$^{a}$& 0.967  & 6     &  10  $\pm$ 2.1  & 77 $\pm$ 22 \\
            &        & 1.45  &  36  $\pm$ 8    & 93 $\pm$ 21 \\
000210$^{b}$& 0.8452 & 2.1   &  $<$32          & $<$80       \\
  000418    & 1.1185 & 6     &  0.5 $\pm$ 3.3  & $<$77       \\
  000911    & 1.0585 & 6     & -0.4 $\pm$ 2.9  & $<$51       \\
  010222    & 1.478  & 6     & -0.7 $\pm$ 2.6  & $<$93       \\
  021211    & 1.006  & 6     &  7.0 $\pm$ 3.5  & $<$120      \\
\hline
\end{tabular}
$^{a}${\ The SFRs reported for GRB\,980703 assume negligible afterglow contribution to our late-time observations.} \\
$^{b}${\ Flux value for GRB\,000210 is from \cite{Greiner+2016}}
\end{minipage}
\end{table}

In every case, including our detection, our observations limit the radio flux to a value well below what had been claimed for the host galaxy in the previous literature (scaling those fluxes to a central frequency of 6 GHz, assuming a standard galaxy spectral index\footnote{We use the convention $F_\nu \propto \nu^{\alpha}$} of $\alpha = -0.75$).   These measurements correspondingly rule out star-formation rates as high as those inferred by earlier works: our limiting SFRs (calculated from our measured flux limits, again assuming $\alpha = -0.75$ and using the method of \citealt{Murphy+2011} and a standard cosmology of $\Omega_\Lambda=0.7$, $\Omega_M=0.3$, $h=0.7$) range from 50--120 $M_\odot$ yr$^{-1}$.  Standard IR-based star-formation rate indicators \citepeg{Calzetti+2013} imply that a star-formation rate of $\sim$100 $M_\odot$ yr$^{-1}$ is required to power a typical threshold ULIRG with $L_{\rm IR} = 10^{12} L_\odot$, so our observations strongly suggest that none of these galaxies are ULIRGs.  The inferred star-formation rate given by our only detection (GRB\,980703), assuming that it is host-dominated (\S \ref{sec:980703}), is in the range of ordinary (non-ultra) LIRGs at approximately 80 $M_\odot$ yr$^{-1}$.  This value exceeds the optical/UV star-formation rate (10--30 $M_\odot$ yr$^{-1}$; \citealt{Djorgovski+1998,Christensen+2004}) only by a factor of a few.

Why did earlier studies infer a luminous, nonfading host galaxy counterpart at the GRB location, whereas our deeper observations rule out such an association?  There are several possibilities, which we consider below.

\subsection{Noise Fluctuations}
\label{sec:notreal}

A quick inspection of Table \ref{tab:prevfluxes} shows that only three of the pre-Swift radio-detected host galaxies have any detections exceeding $>3\sigma$ significance.  For a detection threshold below this level, it is not surprising for some spurious detections to emerge in a large survey (\citealt{Berger+2003} report observations of 17 targets), especially if a small amount of afterglow flux may also be present (next section) or if the RMS noise is slightly underestimated.  This may have been a contributing factor to the marginal radio detection of GRB\,000210 (2.0$\sigma$), as well as for the other GRBs with marginally significant late-time radio excesses mentioned as host candidates in \cite{Berger+2003} which we did not re-observe: GRBs 980329, 000301C, and 000926.\footnote{An additional illustration of this is provided by the case of GRB\,000418.  While we attribute the host detection of Berger et al. (2003) to afterglow contamination (\S \ref{sec:afterglow}), we note that they also reported a fainter secondary source ``G2'' slightly offset from this position with a bridge of emission connecting these sources (their Figure 2).  Neither G2 nor the bridge are visible in our new, much deeper image, suggesting that they were noise fluctuations.}

\begin{figure*}
\centerline{
\includegraphics[width=5in,angle=0]{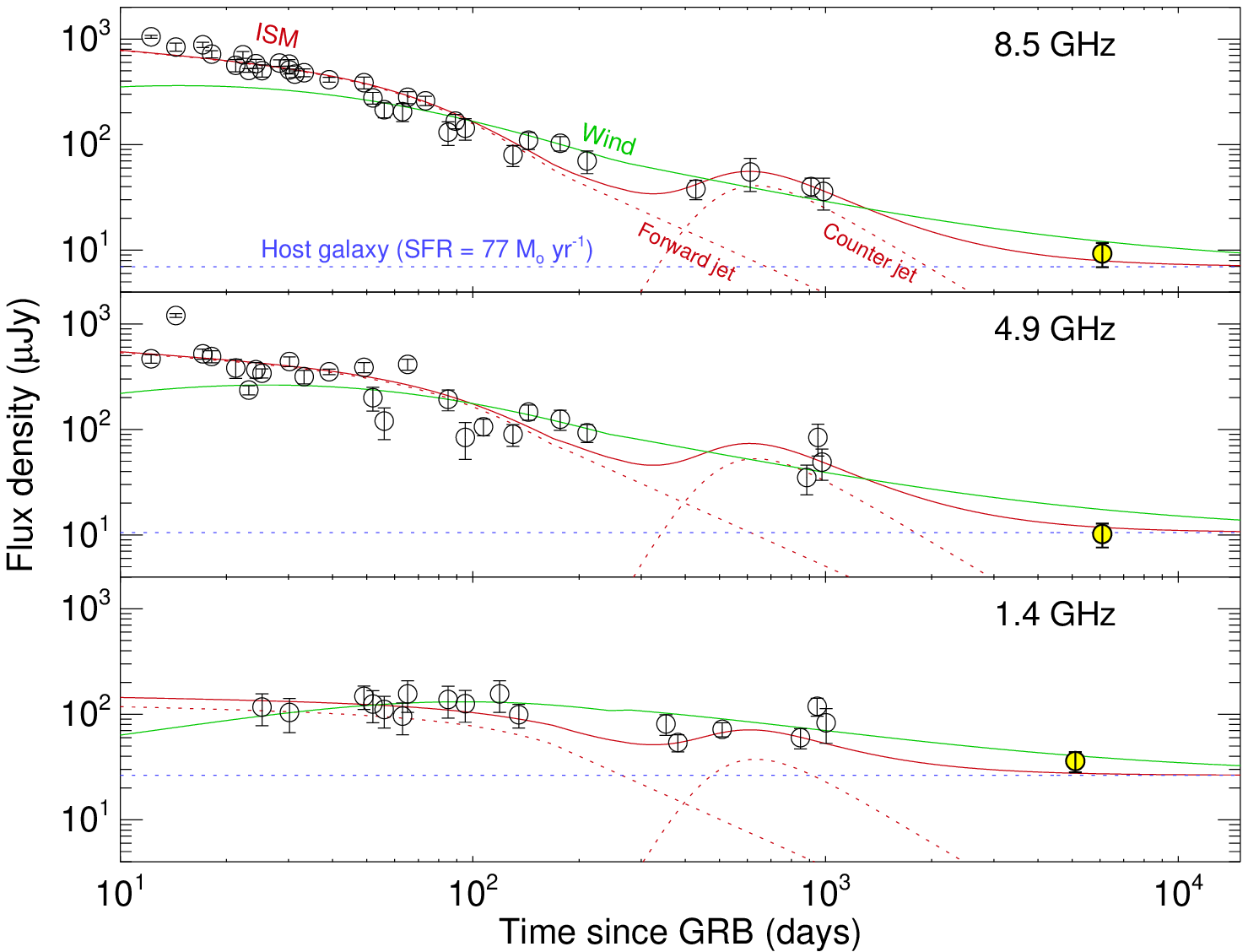}} 
\caption{Multi-frequency radio light curve of GRB 980703, including our new observations (yellow).   Previously, the light curve had seemed to level out in all three bands between 200-1000 days, leading Berger et al.\ (2001) to propose that the host galaxy was dominating the flux.  The new observations show that the source has faded significantly, ruling out this interpretation.   We plot two simple afterglow models:  the solid red curve shows a burst expanding into a constant-density medium and experiencing an early jet break at $\sim$5 days (similar to the original model of \citealt{Frail+2002}), but we associate the late-time flattening with the detection of a \emph{counterjet} 180 degrees off-axis.  (The dotted curves show the two jet components individually.)  The green curve shows a model for a blastwave expanding into an $r^{-2}$ wind with no jet break.  The blue dotted line shows the contribution from a LIRG host galaxy with SFR=77$M_\odot$ yr $^{-1}$.}
\label{fig:grb980703}
\end{figure*}

\subsection{Processing Artefacts}
\label{sec:artifacts}

The detection at the location of GRB\,021211 reported by \cite{Michalowski+2012b} ($330 \pm 31$ $\mu$Jy at 1.4 GHz) is both highly statistically significant and was taken many years after the GRB.  Our deep nondetection only a factor of $\sim$2 in time after this observation conflicts with this measurement (as does the 2~GHz limit from \citealt{Hatsukade+2012}).  To investigate the issue, we downloaded the original L-band observations taken by \cite{Michalowski+2012b} in 2007 from the VLA archive and produced an independent rereduction.  No source is detected at the afterglow position to a flux limit of $<84 \mu$Jy (3$\sigma$).  The reported detection was likely a reduction artefact (Michalowski, priv. comm.)

\subsection{Afterglow Contamination}
\label{sec:afterglow}

Late-time radio light curves for all six pre-Swift GRBs with host radio/submillimetre detections are plotted in Figure \ref{fig:radioag}, combining the afterglow measurements compiled by \cite{Chandra+2012}, the ``host'' measurements of \cite{Berger+2003} at 200--1000 days, and our ultra-late-time observations at $\sim$5000 days.   Data points are color-coded by frequency.

In several cases, the detections at $10^2$--$10^3$ days can be accommodated without difficulty assuming a fairly standard power-law decay (e.g., $t^{-1}$) following the last measurement.  \cite{Berger+2003} did examine the possibility of afterglow contribution to these targets by modeling and extrapolating the multiwavelength light curve, and generally concluded that it is negligible.  However, the \emph{uncertainty} in the model extrapolation was not taken into account in their calculations.  Given the long time baseline, even a slightly slower afterglow decay than the value of their best-fit model could have provided significantly more afterglow flux at the time of their measurements.  In contrast to their results, we find that power-law extrapolation of the radio light curves of GRBs 000418 and 010222 naturally explain the previously-reported ``host'' detections of both targets within 2$\sigma$ uncertainties; these extrapolations are shown (for 8 GHz) in Figure \ref{fig:radioag}.

\subsection{GRB\,980703: Evidence for a counterjet?}
\label{sec:980703}

A similar exercise can be carried out for GRB\,980703: power-law extrapolation of each of its multi-frequency (1, 4, and 8 GHz) light curves individually likewise provides reasonable consistency with the previously-claimed host detections.  However, this event deserves special attention: the dramatically different decay slopes at each frequency indicate that strong spectral evolution was occurring at this time and an unbroken power-law extrapolation is not physically well-motivated.   Specifically, the steep spectral index at early times requires a synchrotron self-absorption break and the spectral evolution requires the passage of the injection break through this band.  Accurately extrapolating the light curve in the presence of these chromatic effects requires a more detailed model incorporating the relevant afterglow physics \citepeg{Sari+1998,Granot+1999}.

The light curve of this event is shown in more detail in Figure \ref{fig:grb980703}.  Two physical models are overplotted: the red curve shows an afterglow expanding into a constant-density medium and experiencing a jet break, while the green curve shows a standard synchrotron afterglow expanding into a wind-stratified ($r^{-2}$) circumburst medium without a jet break.  These models resemble the ones originaly developed by \cite{Berger+2001} and \cite{Frail+2003}, but are re-fit against the radio data after incorporating the revised host-galaxy fluxes and include some further modifications, described below.

The most interesting interpretation is shown by the red curve.  Matching a constant-density model to the early-time data using a single jet leads unavoidably to a large underprediction of the flux at $t>300$ days.  We have therefore added a second component to the model, corresponding to the GRB counterjet.  The counterjet is implemented by a simple empirical prescription as a \cite{Beuermann+1999} broken power law with its rising power-law index, sharpness parameter, and decaying power-law index set to match the numerical light curves of \cite{Zhang+2009}; we use values of 6, 0.6, and -2 respectively.  The spectrum is the same as of the forward jet and its peak time and flux are allowed to vary as free parameters.  We find that the counterjet peaks at 504$\pm$33 days with a peak flux of 41$\pm$6 $\mu$Jy, approximately 5$\times$ the flux of the forward jet at that time.  This is in reasonable agreement with theoretical predictions: \cite{Zhang+2009} predict that the flux ratio of forward and counter jets should be $\sim$6 at peak, and the counterjet peak time should be at $1900 (1+z) E_{\rm iso,53}^{1/3} n_0^{-1/3}$ days.   The original parameters of \cite{Frail+2003} would imply a peak at 1300 days; our revised model does not uniquely solve for $E_{\rm iso}$ or $n_0$ but places the peak between 1300--3400 days.  This is about a factor of 2.5 or more later in time than actually observed, but considering the uncertain physical parameter estimates and approximate treatment the similarity is nevertheless highly suggestive.  

The behavior of the light curve can also be explained without a counterjet.  The alternative green curve (unjetted wind model) underpredicts the early-time flux and overpredicts the late-time flux by a modest factor, but given modeling systematics we cannot confidently rule out this model.  Other afterglow interpretations might also be viable: \cite{Frail+2004} discussed a few different scenarios which could produce a radio light curve that becomes shallower or flattens at late times.  In addition to counterjet and host-galaxy models, their proposed interpretations include transition of the forward jet to the nonrelativistic phase, time-variable microphysical parameters, and late-time energy reinjection.  Further investigation to determine which of these scenarios may apply to GRB\,980703 is beyond the scope of this paper, but we encourage additional modelling of the entire dataset for this burst using modern numerical and analytic methods to provide more insight on this question.

In the discussion so far, we have assumed that our new late-time measurements were dominated by the host galaxy (the blue dashed curve in Figure \ref{fig:grb980703}).  
The best-fit wind model does imply some afterglow contribution to the latest measurements and could indicate that the host may be even fainter than assumed.  Our ISM model (with or without the counterjet) always predicts a minimal afterglow flux at $t>10$ years, although the model extensions referred to in the previous paragraph may be able to provide a longer slow-decay period.  Long-term (multi-decade) radio monitoring and submillimetre observations will be necessary to provide an unambiguous answer.

\subsection{Submillimetre Source Confusion}
\label{sec:detsumm}

Several pre-Swift hosts were also reported to be detected in the submillimetre:  GRBs 000418, 000210 and 010222 have reports of high-significance ($>3\sigma$) detections at 350 GHz.  While we have no new submillimetre observations of our own to report, our radio nondetections call the submillimetre results into question also.  Considering the low-to-moderate redshifts of these three sources ($z$=0.85--1.48), their high submillimetre star-formation rates (560--690 $M_\odot$ yr$^{-1}$; \citealt{Berger+2003}) would imply bright radio emission far in excess of our reported limits.

Afterglow emission is not likely to be a significant source of contamination in the submillimetre band at late times.  Afterglow SEDs are typically quite flat between radio and submillimetre frequencies ($\alpha \sim 1/3$ to $-1/2$; \citealt{Sari+1998}).  Fluxes of 1--5 mJy at 350 GHz would imply radio fluxes of several hundred $\mu$Jy in the radio bands on a similar timescale (1--2 years), which were not observed in any of these cases (Figure \ref{fig:radioag}).  A steeper spectral index could arise if the radio spectrum was self-absorbed, but because the self-absorption break frequency can only decrease with time this would have prevented detection of the radio afterglow at earlier epochs also, contrary to the observations.

It is at least possible that the significance of the submillimetre detections might have been overstated, as may have been the case with some of the earlier radio observations.  To check this, we refer to the independent rereductions of pre-Swift SCUBA data provided by \cite{Michalowski+thesis}.  While the degree of significance of the detections they report are lower than the original values from \cite{Berger+2003}, they confirm statistically-secure detections at both the GRB locations above (2.98 $\pm$ 0.90 mJy at GRB\,000418, and 3.31 $\pm$ 0.60 mJy at GRB\,010222), and a marginal detection of GRB\,000210 (2.81 $\pm$ 1.03 mJy).  Most likely, then, these observations do represent secure detections of astrophysical sources.

The association of these sources with the host galaxy is, however, far from clear.  \cite{Chen+2013} estimate that the density on the sky of sources with $F_{850 \mu \rm{m}} \sim$3 mJy is approximately 3600 per square degree, or 1 per square arcminute.  This implies a covering fraction of 4\% of the sky within the SCUBA beam (FWHM 14$\arcsec$ diameter) around similarly bright sources---quite comparable to the reported detection fraction of 2 out of 26\footnote{A similar calculation was made by \citealt{Tanvir+2004}, who estimated that 1 out of every 20 pre-Swift GRBs would (on average) falsely align with background SCUBA sources.}.  It therefore seems quite plausible that either or both of the two ``secure'' submillimetre detections could represent background (or less likely, foreground) sources.  We do detect several other radio sources within the equivalent JCMT beam of GRB\,010222 in our VLA observations, and the optical and NIR images of these two fields show numerous other sources within the beam in both cases.\footnote{None of the optical sources in the JCMT beam presented by \citealt{Frail+2002} (Figure 3) match the positions of the radio sources we detect, however, and there are no radio sources detected within 7$\arcsec$ of GRB\,000418.  It is probable that both submillimetre sources represent high-$z$ unassociated galaxies with faint radio/optical counterparts.}

We cannot completely rule out that these host galaxies are unusual objects with moderate star-formation rates ($\sim50 M_\odot$ yr$^{-1}$) and unusually low characteristic dust temperatures ($<$30 K; in constrast, typical ULIRG dust temperatures reported by \citealt{Casey2012} range from 35--50 K).   A low dust temperature would shift the SED peak closer to the SCUBA bands, enabling bright submillimetre flux to be observed at 350 GHz even with a modest star-formation rate and weak radio flux.  This would be surprising: compact, luminous, young galaxies of the type typically seen to host GRBs would, if anything, be expected to have \emph{higher} dust temperatures than normal.    Deep observations with a sensitive millimetre interferometer such as ALMA would be required to rule out this possibility unambiguously, but even if the dust temperature is low, the IR luminosity implied by the radio-inferred star-formation rates requires that the hosts would be LIRGs similar to the probable host of GRB\,980703---not ULIRGs as originally claimed by \cite{Berger+2003} and subsequent work.

\section{Conclusions}
\label{sec:conclusions}

In this paper, we have presented VLA re-observations of five pre-Swift GRBs for which luminous host-galaxy counterparts had been reported from radio and/or submillimetre data.  All five counterparts had either disappeared or (in one case) faded to a level far lower than previously claimed, ruling out the presence of an ultraluminous star-forming galaxy at these locations.  A sixth source, GRB\,000210, was not observable to the VLA, but a recent limit from the literature suggests a similar story for this event.

We conclude that most of the preceding radio detections were due to lingering radio afterglow emission or to noise fluctuations or reduction artefacts.  Our results similarly cast doubt on the reported submillimetre detections, suggesting that they originated from source confusion with background SMGs elsewhere in the SCUBA beam.

Our observations offer several lessons relevant to the ongoing Swift era.

\begin{table*}
\begin{minipage}{110mm}
\caption{Swift GRBs localized to ultraluminous host galaxies$^{a}$}
\label{tab:swiftulirgs}
\begin{tabular}{lllllll}
\hline
{GRB} & {$z$} & OA?$^b$ & $A_V$ $^c$   & $M^*$       & {SFR}                 & {Detections$^{c}$}  \\ 
{}    & {}    &         & (mag)        & ($M_\odot$) & ($M_\odot$ yr$^{-1}$) &                    \\ 
\hline
  060814   & 1.920   & IR   & $>2$     & 1.6$\times$10$^{10}$  & 250   &  VLA, SED            \\
  070306   & 1.496   & IR   & $\sim4$  & 5.0$\times$10$^{10}$  & 140   &  VLA, Herschel       \\
  080207   & 2.086   & none & $>3$     & 1.2$\times$10$^{11}$  & 850   &  VLA, Herschel, MIPS \\
\hline
  061121   & 1.314   & yes  & $\sim0$  & 1.5$\times$10$^{10}$  & 160  &   VLA (3$\sigma$) \\
  070521   & 1.1185  & none & $>10$    & 3.1$\times$10$^{10}$  & 800  &   VLA (3$\sigma$) \\
  090404   & $\sim$3?& none & $>1.6$   & 5.5$\times$10$^{10}$  & 1230 &   VLA (4$\sigma$) \\
\hline
\end{tabular}
$^{a}${\ We define an ultraluminous host galaxy as a galaxy with SFR$>100$ $M_\odot$yr$^{-1}$ or $L_{\rm IR} > 10^{12} L_\odot$}.  We regard the ultraluminous nature of the GRB hosts above the horizontal bar as secure on the basis of strong radio detections and confirmation at another frequency.  In the case of GRB060814, SED fitting to the optical and IR photometry also indicates a star-formation rate of $\sim$200 $M_\odot$ yr$^{-1}$.  Those below the bar are less secure due to lower-significance detections and a lack of multiwavelength confirmation.  References: \cite{Svensson+2012,Perley+2013b,Perley+2013a,Perley+2015,Hunt+2014,Greiner+2016}\\
$^{b}${\  Whether or not an optical afterglow was detected for this GRB.} \\ 
$^{c}${\  Line-of-sight extinction towards the GRB as measured from the afterglow.} \\ 
\end{minipage}
\end{table*}

\begin{itemize}  
\item Radio afterglows are truly long-lived objects, peaking on timescales of weeks to months and fading slowly thereafter, potentially remaining detectable for years---a fundamentally different situation from X-ray and optical counterparts which inexoriably are fading after the first day.  While a complication for host searches (next paragraph), this provides advantages for afterglow follow-up.   In the Swift era, the large number of events with near-instant positional notifications has shifted observational emphasis to very early times, with the indirect effect of making long-term dedicated campaigns much less common.   Even so, systematic and complete studies of radio afterglow properties should still be possible for patient observers acquiring data on timescales of months to years.  Interesting physical signatures may be found in such campaigns: our data hints that counterjet emission may have been detected from GRB\,980703 completely accidentally using the pre-upgrade VLA.  If such features are common, detailed study of this behavior in newer bursts should be easily possible with the modern VLA.  Events with similar physical properties as GRB\,980703 (a reddened event with a high inferred circumburst density, leading to chromatic radio evolution and a rapidly fading afterglow due to an early jet break) will be of particular interest for extended radio follow-up campaigns.  (See also the discussion of this point in \citealt{Chandra+2012} and \citealt{Ghirlanda+2013}.)

\item Searches for host galaxies in the radio band require long delays ($>$10 years) to rule out the contribution of a bright and/or late-peaking radio afterglow to any detections.  Radio studies of GRB hosts have enjoyed a resurgence in the past five years \citep{Hatsukade+2012,Perley+2013b,Perley+2015,Berger+2013,Stanway+2014,Nicuesa+2014,Greiner+2016} thanks to the VLA's improved capabilities and similar improvements to other arrays, and while upper limits remain the norm a number of new detections have been reported.  Already, it has become apparent that a few host candidates reported after a delay of only $\sim$1 year were actually long-lived afterglow emission (e.g., GRB\,100621A; \citealt{Greiner+2016}).  Our new results suggest that even delays of several years may not always be enough; re-observations on a timescale of a decade will likely be necessary to avoid the risk of afterglow contamination.  Complementary observations at submillimetre and/or FIR wavelengths represent a crucial test to verify the nature of luminous hosts identified based solely from radio observations, especially in cases where the afterglow evolution is poorly-constrained and/or the radio host detection is of marginal significance.

\item Among the long GRB hosts with radio host detections and ULIRG-scale luminosities that have (so far) survived the test of time (Table \ref{tab:swiftulirgs}), a clear picture is beginning to emerge: they are typically massive and optically-luminous, and the GRBs that occur within them are almost always heavily obscured.   This is exactly what would be expected given our current understanding of the DSG population.  Luminous DSGs are massive galaxies, and the dust covering fraction in front of the youngest stellar population is quite high, concealing the optical afterglow emission from any GRBs which explode within them.  The blue, low-mass ``SMGs'' reported to host several unobscured pre-Swift bursts (by, e.g., \citealt{Berger+2003} and \citealt{Michalowski+2008}) always seemed peculiar from a physical standpoint.  It now appears that galaxies of this type do not exist, or at least do not produce GRBs.  Future radio searches for ultraluminous GRB host galaxies can focus on galaxies known from their optical properties to have stellar masses, optical colors, and/or optical SFRs consistent with known populations of luminous DSGs.
\end{itemize}


\vskip 0.02cm

\section*{Acknowledgments}

D.A.P.\ acknowledges support from a Marie Sklodowska-Curie Individual Fellowship within the Horizon 2020 European Union (EU) Framework Programme for Research and Innovation (H2020-MSCA-IF-2014-660113).  Support for this work was also provided by the National Aeronautics and Space Administration (NASA) through an award issued by JPL/Caltech.  The National Radio Astronomy Observatory is a facility of the National Science Foundation operated under cooperative agreement by Associated Universities, Inc.  We acknowledge useful conversations with and comments from D.~A.~Frail, and from D.~Watson, M.~Micha{\l}owski, and D.~A.~Kann.  We thank the referee for helpful comments.  We also thank A.~Levan for supplying HST imaging for several of the host galaxies.

\bibliographystyle{apj}

\end{document}